\documentclass{osa-article}
\journal{oe}
\usepackage{graphicx}               % Include figure files
\usepackage{lipsum}
\usepackage{bm}                     % bold math
\usepackage{xcolor}                 % I will use for editing 
\usepackage{ulem}                   % for cross-out text
\usepackage{hyperref}               % add hypertext capabilities
\usepackage{physics}
\usepackage{mathtools}
\usepackage[mathlines]{lineno}      % Enable numbering of text and display math
\usepackage{caption}
\usepackage{subcaption}
\usepackage{float}
\usepackage{amsmath}
\usepackage{tcolorbox}
\articletype{Research Article}
\begin{document}

\title{Nonlinear Conversion of Orbital Angular Momentum States of Light }% Force line breaks with \\

\author{Pascal Bassène,\authormark{1,3,*} Finn Buldt,\authormark{1,3}  Nazifa Rumman,\authormark{2} Tianhong Wang,\authormark{1} Phillip Heitert,\authormark{1} and Moussa N'Gom\authormark{1}}

\address{\authormark{1}Department of Physics, Applied Physics and Astronomy, Rensselaer Polytechnic Institute, USA.\\
\authormark{2}Electrical, Computer, and Systems Engineering Department, Rensselaer Polytechnic Institute - Troy, New York 12180-3590.\\
\authormark{3}These authors contributed equally to this work.}
\email{\authormark{*} bassep@rpi.edu} %% email address is required
%%%%%%%%%%%%%%%%%%%%%%%%%%%%%%%%%%%%%%%%%%%%%%%%%%%%%%%%%%%%%%%%%%%%%%%%%%%%%%%%
%%%%%%%%%%%%%%%%%%%%%%%%%%%%%%%%%%%%%%%%%%%%%%%%%%%%%%%%%%%%%%%%%%%%%%%%%%%%%%%%
\begin{abstract}
We investigate the second harmonic generation of light field carrying orbital angular momentum  in bulk  $\chi^{(2)}$ material. We show that due to conservation of energy and momentum, the frequency doubled beam light  has a modified spatial distribution and mode characteristics. Through rigorous phase matching conditions, we demonstrate efficient mode and frequency conversion based on three wave nonlinear optical mixing.
\end{abstract}
%%%%%%%%%%%%%%%%%%%%%%%%%%%%%%%%%%%%%%%%%%%%%%%%%%%%%%%
%%%%%%%%%%%%%%%%%%%%%%%%%%  body  %%%%%%%%%%%%%%%%%%%%%
\section{Introduction}
\begin{figure}[h!]
    \centering
    \includegraphics[width=0.85\textwidth]{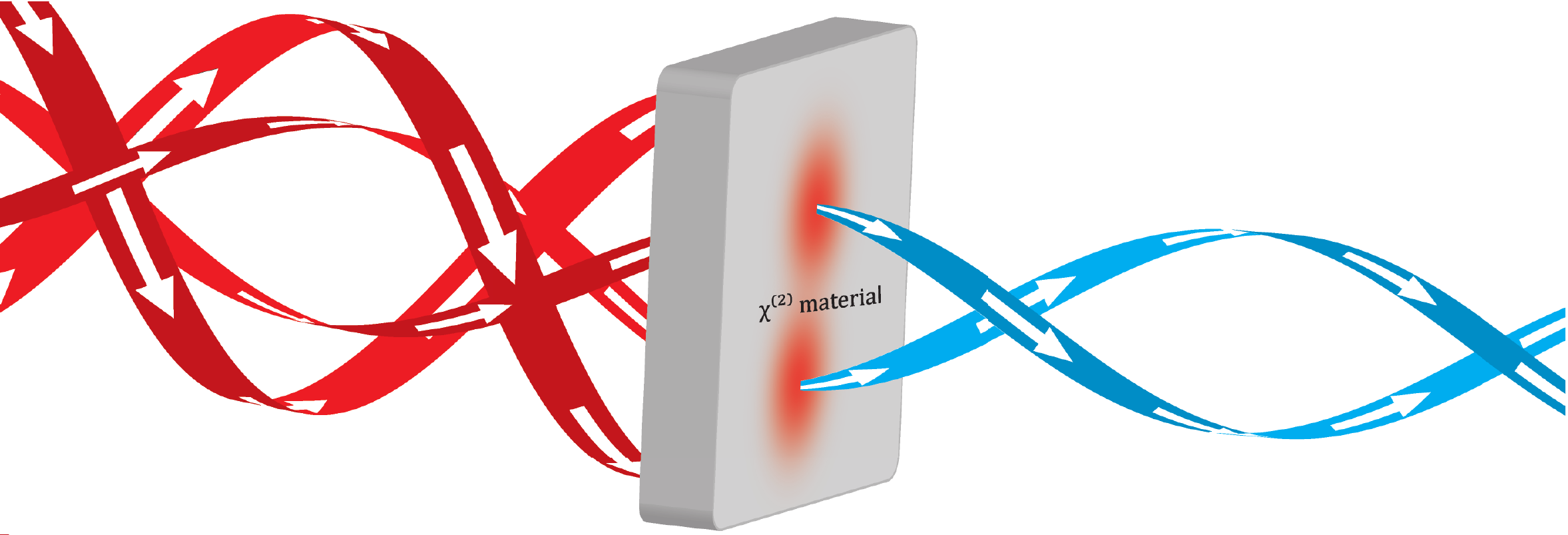}
    \caption{Frequency doubling of an OAM beam in bulk $\chi^{(2)}$ material: in the second-harmonic beam, the number of photons is halved. Consequently the doubling of $\ell$ corresponds to a conservation of the orbital angular momentum within the light field.}
    \label{fig:Illustration}
\end{figure}

Light beams carrying orbital angular momentum (OAM) are purposefully and carefully designed such that their state of polarization is spatially varying as they propagate. This property can lead to new phenomena borne out of exotic light matter interaction that can expand the functionality and enhance the capability of optical systems.\\
OAM lights have an associated azimuthal phase term described by $\exp(i\ell\phi)$, where $\phi$ is the azimuthal angle and $\ell$ is azimuthal mode or the topological charge corresponding to an orbital angular momentum of $\ell\hbar$ per photon\cite{Fontaine2019}. Such phase dependence is characteristic of either Laguerre-Gaussian (LG) or Bessel modes. These beams are primarily formed from interferometric methods that allow the generation of radially and azimuthally polarized beams. Liquid crystal based spatial light modulators (SLM) have introduced flexibility in the generation and manipulation of structured light beams.
Such a flexibility have enabled numerous applications including, but not limited to topological charge transfer in light-matter interactions\cite{He1995}, atomic transitions\cite{Franke-Arnold2017}, spin object detection\cite{Ambrosio2013, Lavery2013}, intrinsic OAM generating laser \cite{Miao2016, Zhang2018}, spin-orbital coupling\cite{Galitski2013}, and quantum information and communication\cite{Mair2001, DaLio2020}.
In quantum communication, there is a need for quantum mode and frequency conversion devices, e.g., based on three or four-wave nonlinear optical mixing \cite{awscha2021}. 
The primary means of achieving this, have been frequency conversion of LG modes that are then transformed into their associated Hermite-Gaussian (HG) modes using a combination of cylindrical lenses\cite{Dholakia1996}.\\
Here, we present a method of frequency doubling and mode conversion  of various OAM beams in bulk  $\chi^{(2)}$ material.  We experimentally demonstrate that the second harmonic beam waist of a LG mode is halved, the amplitude distribution remains, and that the angular momentum per photon is doubled; as illustrated in Fig.\ref{fig:Illustration}. We also generate second harmonic light using HG and Airy-Gaussian (AG) as fundamental input. We demonstrate for the first time, as far as the authors are aware, that HG and AG mode distribution are conserved through frequency doubling. Using meticulous phase matching conditions, we analyze and compare the efficiency of various LG and their corresponding HG modes. 
%%%%%%
\section{\label{Experiementa}Experimental setup} 
\begin{figure}[h!]
\centering
\includegraphics[width=0.85\textwidth]{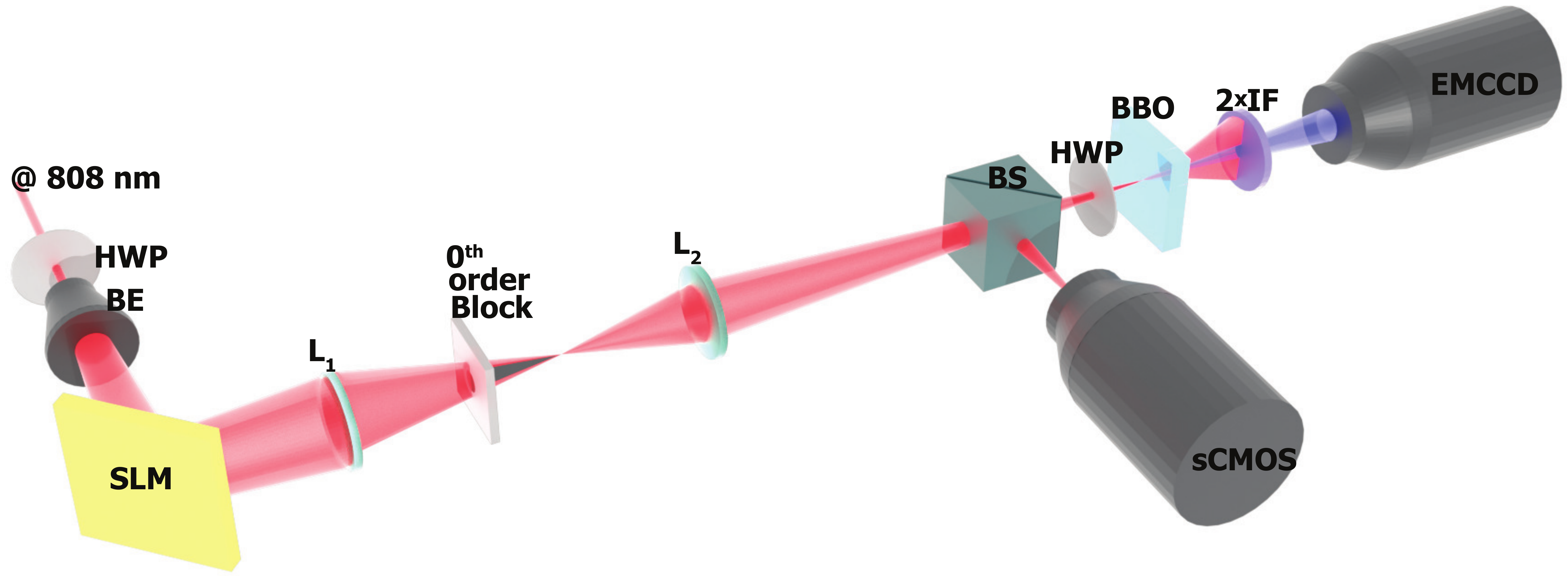} 
\caption{Schematic of the experimental setup:  the beam expander (BE) enlarges the beam to cover the spatial light modulator (SLM). The OAM beam from the SLM is collected by the lens L$_1$. The polarization of the fundamental is controlled by the half wave plate (HWP) before it is focused onto the BBO crystal by L$_2$. The pump signal is characterized using the sCMOS before it is blocked by the filter (IF) and the SHG signal is characterized using the EMCCD}
\label{fig:Setup}
\end{figure}

%%%%%%%
The experimental setup is shown in Fig.\ref{fig:Setup}. The laser source is a mode locked regenerative Ti:sapphire laser with repetition rate of 3kHz,  pulse energy of 1.67 mJ. The laser system can produce  a 38 fs  pulse duration at 808 nm center wavelength. The linearly polarized output of the laser is expanded to cover the entire face of the light modulator. The wavefront manipulator is  a phase-only Santec Liquid crystal based Spatial Light Modulator (SLM-200). Our nonlinear crystal is  a type-I $\beta-$Barium Borate (BBO) 500 micron thick, and optical axis $\theta = 28 ^{\circ}$. The crystal  is affixed to a kinematic mount which allows it to be tilted and rotated. 
\\
We generate the desired phase mask or holographic grating on a computer to display on the SLM. 
All the diffracted orders generated  from the mask are collected by a lens L$_1$ and focused through the BBO by L$_2$. We've designed a homemade aperture to block the 0$^{th}$ order at the focal point of L$_1$. This allows us to discriminate any order of OAM mode generated from the mask. 
The sCMOS camera is used to characterize the shape and intensity of the incident beam (808 nm) on the BBO. The sCMOS and  BBO crystal are equidistant from the  beamsplitter (BS) to ensure that the desired OAM mode is incident on the crystal.
A second HWP is placed before the BBO to further optimize the phase matching conditions. 
The generated second harmonic signal is imaged using an EMCCD placed in the far-field. We use two interference filters centered at 405 $\pm$5 $nm$ to ensure that only the frequency doubled OAM modes are imaged.
%%%%%%%%%
\section{Results and discussion}
%%%%%%%%%%% 
To support our experimental findings, we compare our input (pump) and output (SHG) modes with theoretical results.
The general expression of the amplitude distribution for the Laguerre-Gaussian mode is given by
 %%%%%%%%%%%%%
 %%%%%%%%%%%
 \begin{equation}
\begin{aligned}
E^{(\omega)} \sim  \qty(\frac{\sqrt{2} r}{w})^\ell L^\ell_p \qty(\frac{2 r^2}{w^2})\exp(\frac{ikr^2}{2(z_R^2 + z^2)})  \exp[ - i(\ell + 2p +1)\arctan(\frac{z}{z_R})]  \exp(- i\ell\phi)
\end{aligned}
\label{equ:LG}  
\end{equation}
where $k$ is the wave-number,  $z_R$ is the Rayleigh range,  $w$ is the beam spot or half width of the beam, which depends on z: $w = w_o\sqrt{1 + \frac{z^2}{z_R^2}}$. The  $\exp(- i\ell\phi)$ term describes the helical phase structure of the light, which carries angular momentum $\ell \hbar$ per photon. The integer $p$ and $\ell$ are the polar and azimuthal orders respectively.
%%%%%%%%%%%%%%%%%%%%%%%%%%%%%%%
%%%%
\begin{figure}[h!]
\centering
\includegraphics[width=0.75\textwidth]{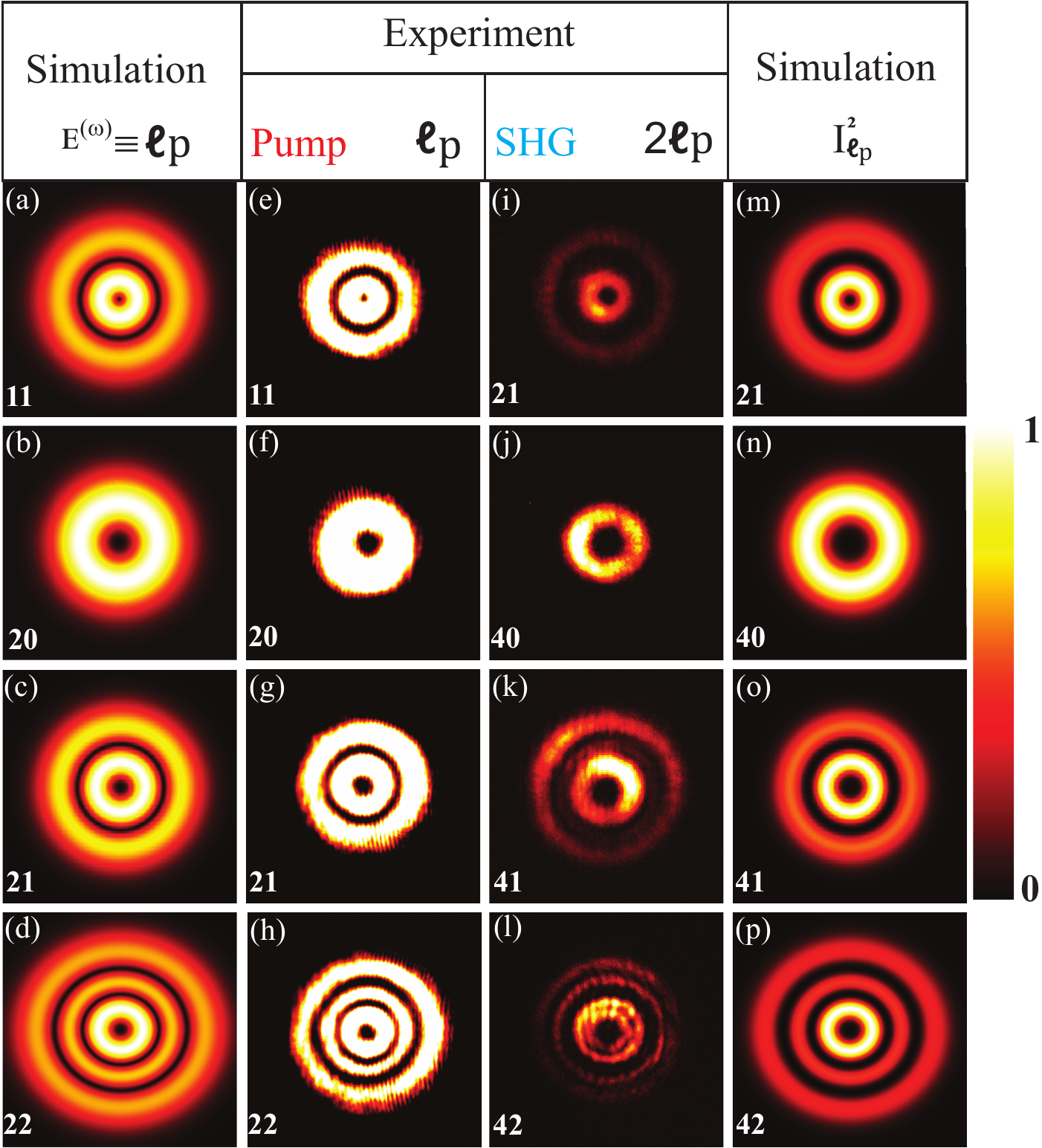}  
\caption{Intensity patterns and profiles of Laguerre-Gaussian beam (LG$_{\ell p}$): Left column shows the numerical simulations of various LG$_{\ell p}$ modes. The center left column shows the corresponding experimental beam profile used to pump (808 nm) the NL crystal. The experimental SH (404 nm) LG modes generated, are displayed in the center right column with azimuthal mode $2 \ell$. The calculated SHG light patterns and profiles are shown in the right most column.}
\label{fig:LGforSHG}
\end{figure}
%%%%%
\\In the two leftmost columns of Fig.\ref{fig:LGforSHG}, we show that our experimentally generated LG$_{\ell p}$ light field accurately reflect the calculated intensity distribution. The various modes generated are used to pump the BBO crystal to produce frequency doubled LG beams. In the center right column of  Fig.\ref{fig:LGforSHG}, we show that the SH signals are also  LG beams with higher order modes. A clear interpretation is provided by Dholakia \textit{et al.}\cite{Dholakia1996}: assuming no absorption and negligible pump depletion; the amplitude distribution  of the SH at frequency $2\omega$ is proportional to the square of the input amplitude distribution at frequency $\omega$; that is $(E^{(\omega)})^2 \propto E^{(2\omega)}$. To confirm this relationship, we plot the intensity distribution of the square of the pump beam:
%%%%
\begin{equation}
 (E^{(\omega)})^2 \sim  \qty(\frac{\sqrt{2} r}{w(z)})^{2\ell} {L^\ell_p \qty(\frac{2 r^2}{{w(z)}^2})}^2 \exp\qty(\frac{-2r^2}{w(z)}) \propto E^{(2\omega)}
\label{equ:Intensity}
\end{equation}\\
%%%%%
In the two rightmost columns of Fig.\ref{fig:LGforSHG}, we display the measured SHG of the LG$_{\ell p}$ modes. We then compare them to the intensity distribution of $(E^{(\omega)})^2$. It is clear that the second harmonic beam experiences a reduction in beam waist, the intensity distribution is proportional to the square of the input beam, and that the angular momentum per photon is doubled: $E^{(\omega)}_\ell \xrightarrow{\chi^{(2)}} E^{(2\omega)}_{2\ell} $.
\\Indeed as energy conservation in the SHG process requires that 2 photons combined their energy $\hbar\omega$ to produce one with twice the energy $2\hbar\omega$. Similarly, photons with orbital angular momentum $\ell\hbar$ combine to yield a resulting photon with OAM $2\ell\hbar$.
%%%%%%%%%%%%%%%%
%%%%%%
\begin{figure}[h!]
\centering
\includegraphics[width=0.75\textwidth]{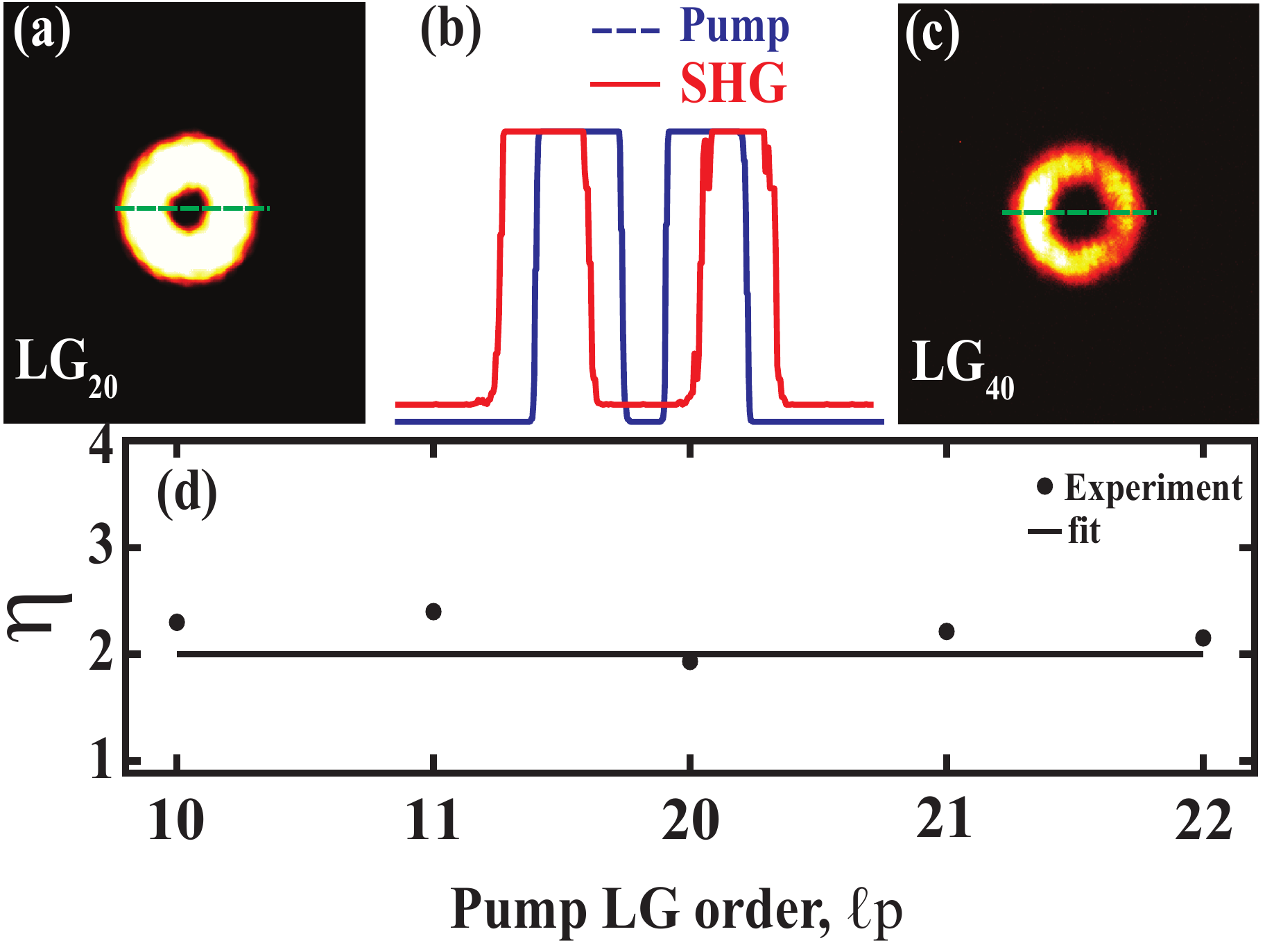}  
\caption{Ratio between SHG and pump inner ring diameter: (a) represents the generated pump LG$_{20}$ mode. (b) shows normalized intensity of the pump and the SHG along the diameter illustrated by the green dash line in (a) and (c). (c) represents the converted mode LG$_{40}$ from the pump. In (d), we report the distribution of the ratio ($\eta$) between the inner diameters of the generated SHG and the corresponding pump for each pump LG$_{\ell p}$ mode, where we show that it is doubled.} 
\label{fig:RatioSHGforPump}
\end{figure}\\
%%%%
It is also clear from Fig.\ref{fig:LGforSHG} that when the radial mode order $p > 0$, the SHG beam propagates in a structurally stable manner even though the squared associated Laguerre polynomial  can no longer be described by a single Laguerre polynomial.\\
Further, the number of photons in the second harmonic beam  is halved, consequently the beam waist will also experience a reduction as illustrated in Fig. \ref{fig:Illustration}. We characterize this by comparing the beam characteristics of the  pump and its corresponding SHG. 
We  show in Fig. \ref{fig:RatioSHGforPump} (a), (b), \& (c) a measure of the  intensity distribution (profile) of the selected LG$_{\ell p}$ modes and its corresponding SHG  modes (LG$_{2\ell p}$). It is shown \cite{Baumann2009} that the radius of the ring, defined by the distance from the center
to points where the intensity is maximum is given by $w^{(2\omega)} = 2w^{(\omega)} \sqrt{\ell/2}$. We verify the beam waist reduction by measuring $\eta =\sqrt{2(\ell'/\ell)}$ where $\ell' = 2 \ell$ is the SHG signal's azimuthal mode. In Fig. \ref{fig:RatioSHGforPump} (d),  the ratio $\eta$ is close to the predicted value with the biggest gap equal to 0.45 for the selected LG$_{11}$ mode. In this way, we have verified that the azimuthal mode $\ell'$ of our converted mode is doubled. Therefore, the momentum and energy are conserved in nonlinear conversion of light carrying OAM. The nonlinear $\qty(\chi^{(2)})$ material in addition to being a frequency converter, can also act as a mode converter.
It is known that LG and HG modes form an orthogonal set.  Mode converters are often  used to convert LG modes into their corresponding HG modes \cite{Dholakia1996}. Here, we directly generate various HG beam using a SLM. Their intensity distribution is compared to their corresponding theoretical expression in Eq. \ref {equ:HG}.  
%%%%%
\begin{equation}
E_{m, n}=\sqrt{\frac{2^{1-m-n}}{\pi m!n!}}\frac{1}{w(z)} H_m\left(\frac{x\sqrt{2}}{w(z)}\right)H_n\left(y\frac{\sqrt{2}}{w(z)}\right)\exp\left(-\frac{r^2}{{w(z)}^2} + ik\frac{r^2}{2R(z)}-i\psi(z)\right) 
\label{equ:HG}  
\end{equation}\\
%%%%%
Here, $\psi(z) = (m + n + 1)\phi_0(z)$ and H$_{m/n}$ is the Hermite polynomial of order $m$ and $n$.
The Electric-field $E_{m,n}$  is a solution of the paraxial wave equation in Cartesian coordinates \cite{Siegman1986, Nomoto2019}. 
%%%%%%%%%%%%
\begin{figure}[h!]
\centering
\includegraphics[width=0.5\textwidth]{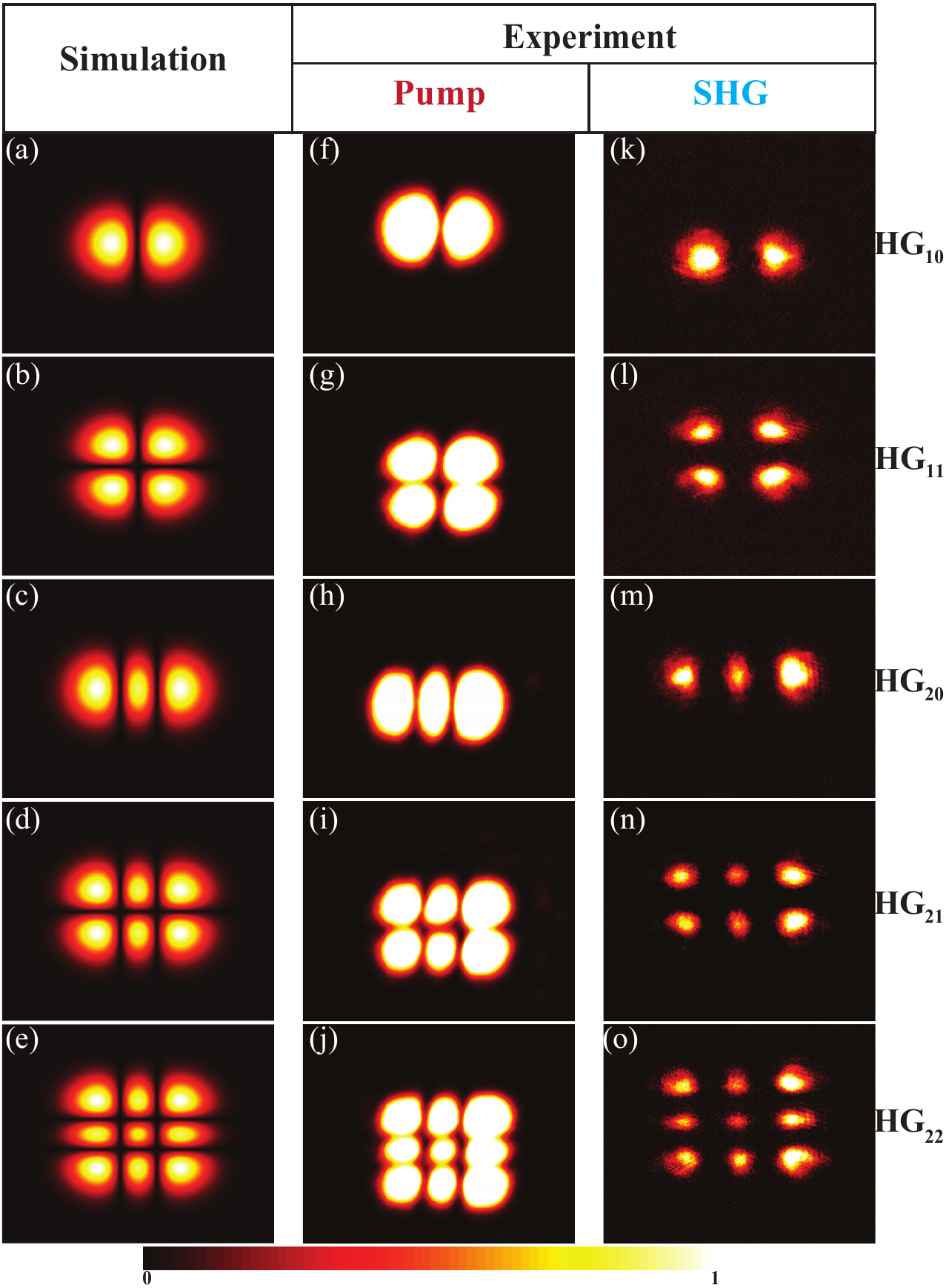}  
\caption{Intensity patterns and profiles of Hermite-Gaussian beam (HG$_{m n}$) and their corresponding SHG modes: Left column shows the numerical simulations of various HG$_{m n}$ modes. The center column shows the corresponding experimental beam profile used to pump the NL crystal. The experimental SH HG modes generated, are displayed in the right column.}
\label{fig:HGforSHG}
\end{figure}
%%%
In Fig.\ref{fig:HGforSHG} we show the intensity distribution of the HG  modes for $m=1,2$ and $n=0,1,2$. The  experimentally generated input (pump) displayed in the center column are compared to the simulated results. The   second harmonic light generated are displayed in the right column . We remark that the generated SHG have the same order ($m$, $n$) that the pump and the simulated modes. Intensity (I$_{\nu}\propto$ E$_{m,n}.$E$^*_{m,n}$) profile of the pump HG modes is slightly dependent to the orders $m$ and $n$ according to equation (\ref{equ:HG}) and compared to LG modes. Therefore, the profile of I$_{2\nu}$ (converted mode) and I$_{\nu}$ are similar even if I$_{\nu}>$ I$_{2\nu}$. The nonlinear conversion of HG modes didn't change the order of the pump modes as shown by Fig.\ref{fig:HGforSHG} (center and right column). As far as the authors know, this is the first time that various orders of HG beam are directly frequency doubled; prior work \cite{Dholakia1996} have often used some variation of a mode converter that involves cylindrical lenses.  Fig.\ref{fig:HGforSHG} also shows that the resulting frequency doubled HG beam have the same spatial distribution as the input HG beam. 

\begin{figure}[hbt!]
\centering
 \includegraphics[width=0.75\textwidth]{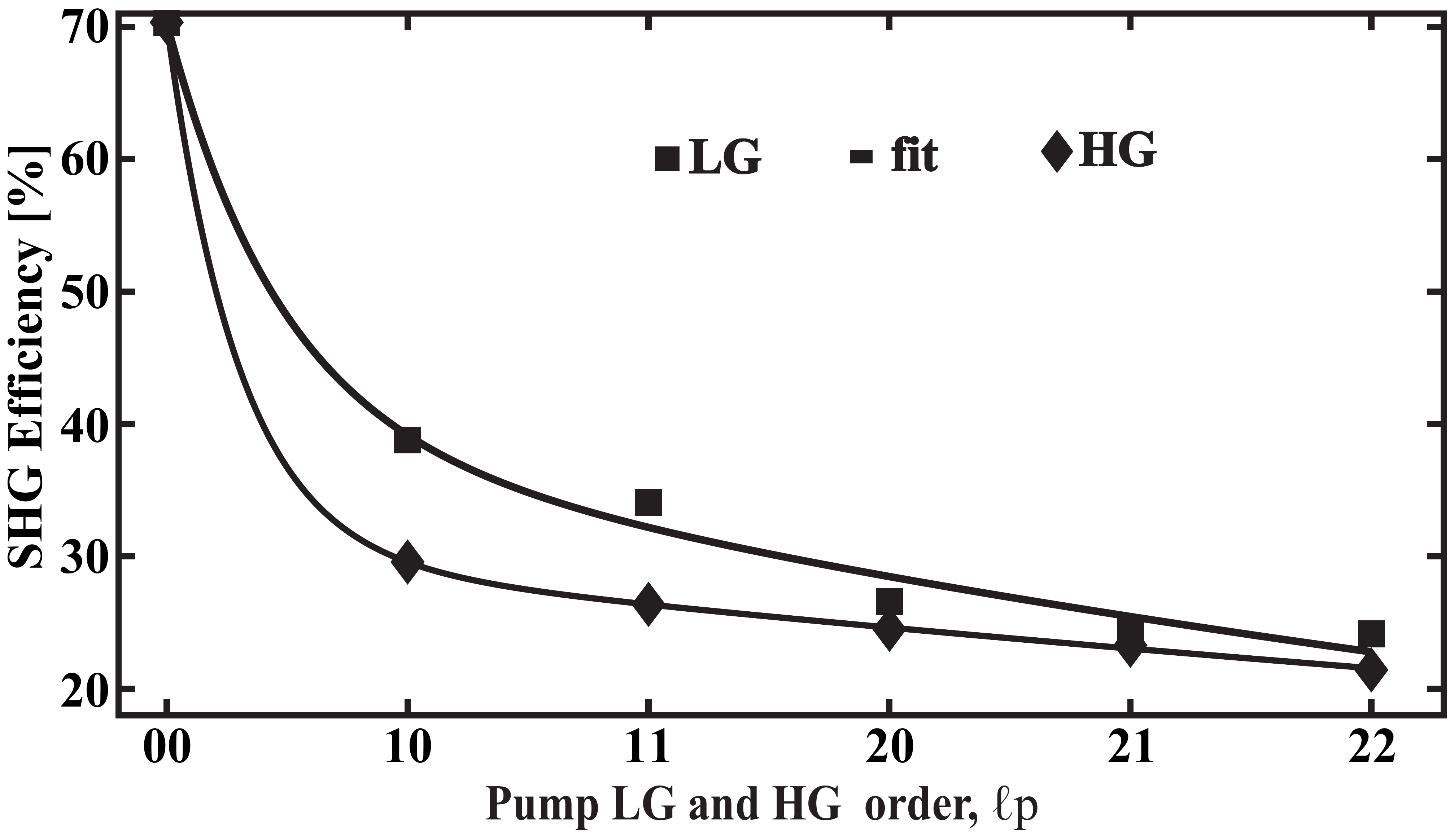} 
 \caption{Second harmonic conversion efficiency of LG$_{\ell p}$ and HG$_{m n}$  beam. The same phase matching conditions are used for both LG and HG modes. The efficiency track well specially for higher order modes}
 \label{fig:Efficiency}
\end{figure}

The SLM in our experimental setup allows for any Gaussian beam to be `transformed' into any structured beam that can be use as a fundamental beam for SHG. For example, we generated  Airy-Gauss (AG) beam  described by:
%%%%%%%%
\begin{equation}
\begin{aligned}
AG = {} exp\left(-\frac{x^2 + y^2}{{\omega_0}^2} + \frac{ax +by}{\omega_0}\right) Ai\left(\frac{\beta y}{\omega_0}\right)Ai\left(\frac{\alpha x}{\omega_0}\right)
\end{aligned}
\label{equ:AG}  
\end{equation}\\
%%%%%%%
where, $a$ and $b$ are truncated factors of the Airy beam and $\alpha$ and $\beta$ arbitrary transverse scaling factors in the $x$ and $y$ directions \cite{Siegman1986, Zhu2020}. \\
%%%%%%%%%%
\begin{figure}[h!]
\centering
\includegraphics[width=0.75\textwidth]{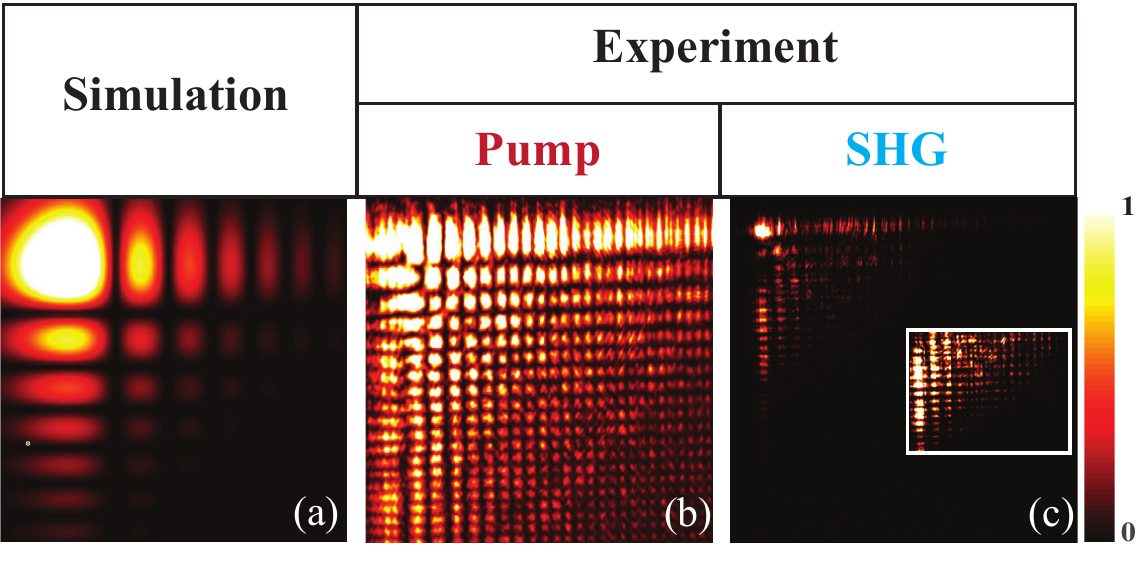} 
\caption{Intensity patterns and profiles of Airy-Gaussian beam (AG): (a) shows the numerical simulation of the AG modes. (b) shows the corresponding experimental beam profile used to pump the NL crystal. (c) the experimentally generated second harmonic AG modes.}
\label{fig:AGforSHG}
\end{figure}\\
%%%%%%%%%%%
Fig.\ref{fig:AGforSHG} shows theoretical and experimental intensity profile of AG modes generated by the SLM and frequency doubling by the BBO. The generated pump Fig.\ref{fig:AGforSHG}(b) shows a multimodes intensity decreasing from the zero order position. Frequency doubling modes intensity decrease exponentially according to equation (\ref{equ:AG}). The inset in Fig.\ref{fig:AGforSHG}(c) shows the modes profile without the zero order brightness. The nonlinear conversion of AG beam is relevant if all its modes can be down-converted and generate entangled photons.
%%%%%%%%
\\
We next characterize the  efficiency of the nonlinear conversion of the LG$_{\ell p}$ and HG$_{m n}$ beams produced by the SLM.  
The SHG conversion efficiency for different order LG$_{\ell p}$ and HG$_{m n}$ modes is plotted in Fig.\ref{fig:Efficiency}. Since the LG and HG  modes form a complete orthonormal set, it is  reasonable to assume that the optimum phase matching conditions is the same for both. It is noticeable in Fig.\ref{fig:Efficiency} that the conversion efficiency for the lower order LG modes are higher than the HG modes. But the efficiency for both modes is similar for higher order modes. The exponential decrease of SHG efficiency with the pump order is related to the decrease of the pump intensity with the mode order. We note that using stringent phase matching method, we are able to obtain efficiency $\sim$70\% for a gaussian input. Fig.\ref{fig:Efficiency} show that the conversion efficiency of the higher  LG and HG modes are similar and follow the same trend.
%%%%%
\\
%%%%%%%%%%%%
%%%%%%%%%%%%
\section{Conclusion}
We have experimentally demonstrated a general approach to  second harmonic generation using light fields carrying orbital angular momentum. We show that the frequency doubled Laguerre-Gaussian signal have reduced beam waist, and their azimuthal mode number $\ell$  is double that of the input LG light, and that their radial mode $p$ is conserved. We have also shown that the second harmonic signal from a Hermit-Gaussian input is also a HG with reduced spatial distribution. 
 Our work adds to the body of knowledge associated  with OAM and SHG research; we've shown that this method can be used as a mode and frequency converter and the flexibility offered by the use of a spatial light modulator allows  for the generation of a variety of structured beams. We believe this work will have fundamental contributions and add practicality in photonics, quantum science, and the physics for OAM devices.
%%%%%%%%%%%%%%%%
%%%%%%%%%%%%%%%%
\begin{backmatter}
\bmsection{Funding} This work is supported in part by the National Geospatial Intelligent Agency grant \# HM04762010012
\end{backmatter}
%%%%%%%%%%%%%%%%%%%%%%%%%%%%%
%%%%%%%%%%%%%%%%%%%%%%%%%%%%%%

\end{document}